\documentclass[11pt]{article}
\usepackage{amsmath,amssymb}
\usepackage{graphicx,color}

\numberwithin{equation}{section}

\def\({\left(}
\def\){\right)}

\newcommand{\de}{\partial}
\newcommand{\be}{\begin{equation}}
\newcommand{\ba}{\begin{eqnarray}}
\newcommand{\ea}{\end{eqnarray}}
\newcommand{\ee}{\end{equation}}

\newcommand{\f}{\frac}
\newcommand{\s}{\sqrt}

\newcommand{\ti}{\tilde}
\newcommand{\ap}{\alpha}

\newcommand{\ddd}{\cdot\cdot\cdot}
\newcommand{\no}{\nonumber \\}

\newcommand{\ep}{\epsilon}

 \def\de{\partial}

 \def\f {\frac}
 \def\ti{\tilde}
 \def\ap{\alpha}

 \def\ddd{\cdot\cdot\cdot}
 \def\no{\nonumber \\}

 \def\ep{\epsilon}

\begin{document}

\begin{titlepage}
\thispagestyle{empty}

\begin{flushright}
YITP-12-42\\
IPMU12-0087\\
\end{flushright}

\vspace{.4cm}
\begin{center}
\noindent{\Large \textbf{Central Charges for BCFTs and Holography}}\\
\vspace{2cm}

Masahiro Nozaki $^{a}$\footnote{e-mail:mnozaki@yukawa.kyoto-u.ac.jp},
Tadashi Takayanagi $^{a,b}$\footnote{e-mail:takayana@yukawa.kyoto-u.ac.jp},
and Tomonori Ugajin $^{a,b}$\footnote{e-mail:tomonori.ugajin@ipmu.jp}

\vspace{1cm}
  {\it
 $^{a}$Yukawa Institute for Theoretical Physics,
Kyoto University, \\
Kitashirakawa Oiwakecho, Sakyo-ku, Kyoto 606-8502, Japan\\
\vspace{0.2cm}
 $^{b}$Kavli Institute for the Physics and Mathematics of the Universe,\\
University of Tokyo, Kashiwa, Chiba 277-8582, Japan\\
 }

\vskip 2em
\end{center}

\vspace{.5cm}
\begin{abstract}
In this paper, we study the logarithmic terms in the partition functions of CFTs with boundaries (BCFTs). In three dimensions, their coefficients give the boundary central charges, which are conjectured to be monotonically decreasing functions under the RG flows. We present a few supporting evidences
from field theory calculations. In two dimensions, we give a holographic construction (AdS/BCFT) for an arbitrary shape of boundary and calculate its logarithmic term as well as boundary energy momentum tensors, confirming its consistency with the Weyl anomaly. Moreover, we give perturbative solutions of gravity duals for the three dimensional BCFTs with any shapes of boundaries. We find that the standard Fefferman-Graham expansion
breaks down for generic choices of BCFT boundaries.
\end{abstract}

\end{titlepage}

\newpage

\begin{scriptsize}
\tableofcontents
\end{scriptsize}

\newpage

\section{Introduction}

The AdS/CFT correspondence \cite{Maldacena,GKP} offers us a non-perturbative framework which
relates gravity theories to conformal field theories (CFTs) in remarkable ways. Usually, the AdS/CFT is considered for a CFT defined on a manifold without any boundaries. However, the properties of quantum field theories with boundaries are also very intriguing. They are sensitive to their boundary conditions and thus a large variety of possible theories are possible. Also in condensed matter physics, field theories with boundaries appear in important systems such as the quantum Hall effects or topological insulators.

Recently, an extension of AdS/CFT to the cases where the CFT is defined on a manifold with boundaries (AdS/BCFT) has been proposed in \cite{Ta}. In specific examples, the same construction has been already mentioned in \cite{KaRa}. In the paper  \cite{Fujita:2011fp}, the partition functions in AdS/BCFT have been computed and a holographic proof for the g-theorem \cite{AfLu,FrKo} has been given with a proposal of its higher dimensional generalization. A string theory embedding of the AdS/BCFT was also given in \cite{Fujita:2011fp}.
The AdS/BCFT has been analyzed in a three dimensional gravity with higher curvatures in \cite{Kwon:2012tp}.
In \cite{Fujita:2012fp}, the AdS/BCFT is employed for a holographic construction of the quantum Hall effect and its edges states. See also \cite{Alishahiha:2011rg,Setare:2011ey} for other developments. A short review can be found in section 4 of \cite{Takayanagi:2012kg}. For other approaches to gravity duals of CFTs with boundaries, refer to \cite{AMR,ABBS,CDGG,GSW}.

The purpose of this paper is to explore the construction and properties of AdS/BCFT. We especially
focus on the logarithmically divergent terms in the Euclidean partition function of BCFTs.
In even dimensional BCFTs, the coefficients of the log terms are related to the Weyl anomaly and thus the central charges. In odd dimensions, on the other hand, these coefficients lead to new quantities called boundary central charges $c_{bdy}$. The logarithmic term in AdS$_4/$BCFT$_3$ is especially intriguing. The holographic analysis in \cite{Fujita:2011fp} shows that the corresponding boundary central charge, extended to the non-conformal
theories as a c-function, gets monotonically decreased under the RG flow:
\be
\f{dc_{bdy}(r)}{dr}\leq 0,
\ee
where $r$ is a length scale of the BCFT. This can be regarded as a higher dimensional analogue of the
g-theorem \cite{AfLu,FrKo}. A part of main results in this paper is to give a few modest evidences for this property from quantum field theoretic calculations, based on a perturbation theory and an explicit example. Finally, we conjecture this c-theorem in arbitrary odd dimensional BCFTs.

So far, the examples of AdS/BCFT have been limited to the cases where boundaries of BCFTs are either
hyperplanes or round spheres. Therefore we would like to
consider the examples where the boundaries are
general curved surfaces. We will show that the coefficient of the logarithmic term in the AdS$_3/$BCFT$_2$ setup is topological (proportional to the Euler number) and thus does not change under smooth deformations of the boundaries. We will also independently confirm this by calculating the energy momentum tensor at the boundary.
Moreover, we will find that in higher dimensional setups, the construction of solutions based on the standard Fefferman-Graham expansion does not work and instead we will construct perturbative solutions by using the hyperbolic foliation of
the AdS space for AdS$_4/$BCFT$_3$.

This paper is organized as follows:
In section 2, we will first give a brief overview of the AdS/BCFT construction. Later we provide a careful treatment of the new codimension two boundary term, which has been neglected previously and calculate the energy momentum tensor localized at the boundary $P$.
In section 3, we will examine AdS duals of two dimensional BCFTs with general shape of boundaries based on the standard Fefferman-Graham expansion and calculate the logarithmic term. In section 4, we present perturbative solutions for AdS$_4/$BCFT$_3$ with general shape of boundaries. In section 5, we argue the higher dimensional g-theorem in terms of boundary central charges and give some evidences. Only this section is purely field theoretical and does not employ the holography. In section 6, we summarize our conclusions and discuss future problems.

\section{AdS/BCFT Formulation and Energy Momentum Tensor}

Here we will first give a brief summary of the AdS/BCFT i.e. a holographic dual of CFT defined on a manifold $M$ with a boundary $\de M(\equiv P)$ \cite{Ta}. Later we will provide a careful treatment of the new codimension two boundary term, which has been neglected previously and we will calculate the energy momentum tensor localized at the boundary $P$.

In AdS/CFT \cite{Maldacena}, a $d+1$ dimensional AdS space (AdS$_{d+1}$) is dual to a $d$ dimensional CFT. The geometrical $SO(d,2)$ symmetry of AdS is equivalent to the
conformal symmetry of the CFT. When we put a $d-1$ dimensional boundary to a $d$ dimensional CFT such that the presence of the boundary breaks $SO(2,d)$ into $SO(2,d-1)$, this is called a boundary conformal
field theory (BCFT) \cite{Cbcft}.

\begin{figure}
   \begin{center}
     \includegraphics[height=4cm]{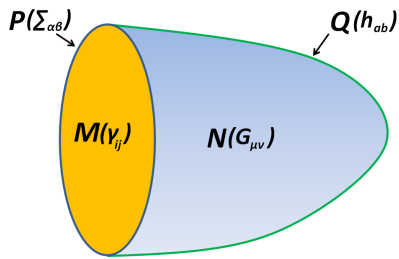}
   \end{center}
   \caption{A schematic setup of AdS/BCFT. The CFT lives on $M$, which has the
   boundary $P\equiv \de M$. Its gravity dual is denoted by $N$ and its asymptotically
   AdS boundary is $M$. The boundary $P$ is extended into the bulk AdS, which
   constitutes the boundary $Q$.}\label{fig:setup}
\end{figure}

The construction of AdS/BCFT goes as follows\footnote{One may think this construction of
AdS/BCFT looks similar to the holographic entanglement entropy \cite{RT}. However, they are
crucially different because of the following reasons. To calculate the holographic entanglement entropy, we pick up a codimension two minimal area surface and this exists in any asymptotically AdS backgrounds. However, the surface $Q$ in the AdS/BCFT is codimension one and has more constraints due to the boundary condition (\ref{eqbein}) and there is no solution in a genetic asymptotically AdS backgrounds. Therefore the boundary
$Q$ backreacts with the bulk spacetime and changes its metric so that the boundary condition is satisfied. Mathematically, the minimal surface condition is equivalent to the vanishing of the trace of the extrinsic curvature i.e. $K=0$, while (\ref{eqbein}) constrains each component of $K_{ab}$.} (refer to Fig.\ref{fig:setup}). The holographic dual of a BCFT (boundary conformal field theory) on a $d$ dimensional manifold $M$
is defined as a gravity on a $d+1$ dimensional spacetime $N$. $N$ is an asymptotically
AdS space and its AdS boundary coincides with $M$. We assume that $M$ has a boundary $\de M$ and in the gravity dual, $\de M$ is extended to a $d$ dimensional manifold
$Q$ such that $\de N=M\cup Q$. To respect the $SO(2,d-1)$ symmetry of BCFT, $N$ should be foliated by $AdS_{d}$ slices. We can also generalize this construction into the
non-conformal cases by relaxing the $SO(2,d-1)$ symmetry. This is the basic setup of AdS/BCFT.  Next we need to impose an appropriate boundary condition on $Q$.

\subsection{Neumann Boundary Condition}

In the standard AdS/CFT, we impose the Dirichlet boundary condition at the boundary of AdS and therefore we require the Dirichlet boundary condition on $M$. On the other hand, we impose a Neumann boundary condition on $Q$ \cite{Ta}. The reason for this is that this boundary should be dynamical from the viewpoint of holography and there is no
natural definite metric on $Q$ specified from the data in the CFT side. Also this
can be naturally derived in the orientfold construction in string theory as in the
example discussed in \cite{Fujita:2011fp}.

To make the variational problem sensible, we need to add the Gibbons-Hawking boundary term \cite{GHterm} on the boundaries $M$ and $Q$ to the Einstein-Hilbert action:
\be
I=\f{1}{16\pi G_N}\int_{N}\s{-G}(R-2\Lambda)+\f{1}{8\pi G_N}\int_{Q}\s{-h}K
+\f{1}{8\pi G_N}\int_{M}\s{-\gamma}K. \label{act}
\ee
The metric of $N$ is defined by $G_{\mu\nu}$, where the index $\mu$ runs the $d+1$ coordinates; the induced metric on $Q$ and $M$ are denoted by $h_{ab}$
and $\gamma_{ij}$, respectively, while
$a$ and $i$ run the $d$ coordinates. For later convenience, we also define
the induced metric on $P(=\de M=\de Q)$ to be $\Sigma_{\ap\beta}$. We summarize these conventions
in Fig.\ref{manifolds}.

$K=h^{ab}K_{ab}$
(or $K=\gamma^{ij}K_{ij}$) is the trace of extrinsic curvature on $Q$ (or $M$).
The extrinsic curvature $K_{ab}$ is defined by
\be
K_{ab}=\nabla_a n_b,
\ee
where $n$ is the unit vector normal to $Q$ and here we implicitly assume a projection onto $Q$ from $N$.
For example, in the Gaussian normal coordinate system, we have the following metric
\be
ds^2=d\eta^2+h_{ab}(\eta,u)du^adu^b, \label{Gauss}
\ee
where $Q$ is situated at $\eta=\eta_*$ and $N$ is given by $\eta\leq \eta_*$. In this setup,
we can explicitly calculate the extrinsic curvature as
\be
K_{ab}=\f{1}{2}\f{\de h_{ab}(\eta_*,u)}{\de \eta}.\label{formulak}
\ee

Now let us consider the variation of metric in the above action. After a partial integration, we find
\be
\delta I=\f{1}{16\pi G_N}\int_{Q}\s{-h}(K_{ab}-Kh_{ab})\delta h^{ab}
+\f{1}{16\pi G_N}\int_{M}\s{-\gamma}
(K_{ij}-K\gamma_{ij})\delta \gamma^{ij}. \label{variation}
\ee

Notice that the terms which involve the derivatives of $\delta h_{ab}$ and $\delta \gamma_{ij}$ cancel out thanks to the boundary term. It is clear that the variation on
$Q$ is vanishing if we impose either the Dirichlet boundary condition $\delta h^{ab}=0$ or the Neumann boundary condition
\be
K_{ab}-h_{ab}K=0. \label{boundary}
\ee
As we mentioned, we choose the Neumann condition (\ref{boundary}) on $Q$, while we
do the Dirichlet one on $M$.

It is also possible to add some matter fields localized on $Q$ and consider a generalized action by adding
\be
I_{Q}=\int \s{-h}L_{Q}.
\ee
This modifies (\ref{boundary}) into
\be
K_{ab}-h_{ab}K=8\pi G_N T^{Q}_{ab}, \label{bein}
\ee
where we defined the energy momentum tensor on $Q$
\be
T^{Q}_{ab}=-\f{2}{\s{-h}}\f{\delta I_Q }{\delta h^{ab}}. \label{matbc}
\ee

In this paper we only consider the case where the boundary matter lagrangian
$L_Q$ is simply a constant $L_{Q}=-\f{T}{8\pi G_N}$. The constant $T$ can be interpreted as the tension of the `brane' $Q$.
The boundary condition (\ref{bein}) for this system reads
\be
K_{ab}=(K-T)h_{ab}.\label{eqbein}
\ee
By taking its trace, we obtain
\be
K=\f{d}{d-1}T.
\ee

\begin{figure}
\begin{center}
\begin{tabular}{|c|c|c|c|}
  \hline
  Manifold & Dimension & Metric & Relations \\ \hline
  $N$ & $d+1$ &  $G_{\mu\nu}$ & $G_{ij}=\f{g_{ij}}{\rho}$ \\ \hline
  $Q$ & $d$ & $h_{ab}$ &  \\ \hline
  $M$ & $d$ &  $\gamma_{ij}$ & $\gamma_{ij}=G_{ij}|_{\rho=\ep}$\\ \hline
  $P$ & $d-1$ & $\Sigma_{\ap\beta}$ & $\Sigma_{\ap\beta}=\f{\sigma_{\ap\beta}}{\rho}$ \\
  \hline
\end{tabular}
\end{center}
\caption{A summary of notations on the manifolds and their metrics in this paper. Notice that
$N$ is the original spacetime where the gravity dual lives. $M$ is its AdS boundary and
$Q$ is the other part of the boundary of $N$. $P$ is defined by $P=\de M=\de Q$.} \label{manifolds}
\end{figure}

The Euclidean formalism of AdS/BCFT is also useful especially for the evaluations of the partition functions and we will mainly employ this formalism in the rest of this paper. In the Euclidean formulation, the gravity action (\ref{act}) in the Lorentzian signature is now replaced by \be I_E=-\f{1}{16\pi
G_N}\int_{N}\s{g}(R-2\Lambda)-\f{1}{8\pi
G_N}\int_{Q}\s{h}(K-T)-\f{1}{8\pi
G_N}\int_{M}\s{\gamma}K,\label{acte} \ee where we added the tension $T$ contribution on $Q$.
Note that in the actual calculations we need to add the counter terms to (\ref{acte}) as in the
standard holographic renormalization of AdS/CFT \cite{HeSk,BaKr,Sk}.

\subsection{Simple Examples} \label{examples}

Here we briefly review the basic examples of AdS/BCFT, which are useful in our later
arguments. We only consider the $d+1$ dimensional pure gravity theory. The first example is the BCFT on a half plane \cite{Ta,Fujita:2011fp}. The metric of AdS$_{d+1}$
with the radius $L$ can be rewritten as follows:
\be
ds^2=d\eta^2+\f{\cosh^2(\eta/L)}{z^2}
(dz^2+d\vec{x}^2),\label{metads}
\ee
where $\vec{x}\in R^{d-1}$. If we assume that $\eta$ takes all values from $-\infty$ to
$\infty$, then (\ref{metads}) is equivalent to the AdS$_{d+1}$. To
see this, define new coordinates $w$ and $\xi$ by
\be
w=\f{z}{\cosh(\eta/L)},\ \ \ \xi=z\tanh(\eta/L). \label{core}
\ee
In this new coordinate system, (\ref{metads}) indeed coincides with the Poincare metric:
\be
ds^2=L^2\left(\f{dw^2+d\xi^2+d\vec{x}^2}{w^2}\right). \label{pads}
\ee
Note that the cosmological
constant $\Lambda$ is related to the AdS radius $L$ by
$\Lambda=-\f{(d-1)d}{2L^2}$.

To realize a gravity dual of BCFT, we will put the boundary $Q$ at
$\eta=\eta_*$ and this means that we restrict the spacetime to the
region $-\infty<\eta<\eta_*$. The extrinsic curvature on $Q$ reads
\be K_{ab}=\f{1}{L}\tanh\left(\f{\eta_*}{L}\right)h_{ab}. \ee By imposing
the boundary condition (\ref{eqbein}), we find the relation
\be
T=\f{d-1}{L}\tanh\f{\eta_*}{L}.\label{tension} \ee
In this system, the AdS boundary $M$ is given by the half place defined by
$\xi\leq 0$.

We can perform the conformal transformation so that the boundary $P=\de M$ is
mapped from the hyperplane to a round sphere \cite{Ta,Fujita:2011fp}. The holographic dual of a BCFT on a round ball with radius $r_B$ is given by
the following region in the Poincare AdS$_{d+1}$ (\ref{pads});
\be
\xi^2+\vec{x}^2+\left(w-r_B\sinh(\eta_*/L)\right)^2-r_B^2\cosh^2(\eta_*/L)\leq
0. \label{diskh}
\ee

\subsection{Codimension Two Boundary Term}

Moreover, strictly speaking, we need to add the boundary term on $P(=\de M=\de Q)$ to the gravity action $I_E$ in (\ref{acte}). This is because $Q$ and $M$ are joined non-smoothly on $P$ with cusp like singularities. In such a case, we need to add the following boundary term \cite{Hay}
\be
I^{(bdy)}_{E}=\f{1}{8\pi G_N}\int_{P}\s{\Sigma}\cdot (2\theta-\pi), \label{boundr}
\ee
where $2\theta$ is the angle between $Q$ and $M$ at $P$ (the angle is measured from inside of $N$). See the appendix \ref{ap:cusp} for an elementary derivation of (\ref{boundr}). $\Sigma_{\ap\beta}$ is the induced metric on $P$. In other words, if we define $n_M$ and $n_Q$ are unit normal vectors toward the outside of the gravity dual $N$, then we have
\be
n_M\cdot n_Q=\cos (\pi-2\theta).
\ee
Therefore the correct gravity action of AdS/CFT is given by (\ref{acte}) plus (\ref{boundr})
i.e.
\be
I^{(tot)}_{E}=I_E+I^{(bdy)}_{E}+I^{(c.t.)}_E,  \label{totac}
\ee
where we also added the counter terms $I^{(c.t.)}_E$
so that total action $I^{(tot)}_{E}$ becomes finite.

Below we would like to examine how the calculations of Euclidean partition functions are affected by this codimension two boundary term (\ref{boundr}). We concentrate on the example of the round
disk partition function in AdS$_3/$BCFT$_2$. The holographic dual of a BCFT defined on a round disk with the radius $r_B$  is given by the gravity on the manifold (\ref{diskh}) inside the Poincare AdS$_3$. The main part $I_E$ has been already calculated in \cite{Ta,Fujita:2011fp}. In the
presence of the new boundary term $I^{(bdy)}_{E}$, the final result reads
\ba
I_E\!+\!I^{(bdy)}_{E}\!
=\!\f{L}{4G_N}\Biggl[\!-\!\f{r_B^2}{2\ep^2}\!-\!\f{r_B}{\ep}
\left(\sinh\f{\eta_*}{L}\!+\!\arccos LT\right)
\!+\!\log \f{\ep}{r_B}\!-\!\f{1}{2}\!-\!\arccos LT\cdot \sinh\f{\eta_*}{L}\!-\!\f{\eta_*}{L}\Biggr], \label{dive}
\ea
where $\ep$ is the UV cut off, set by $z>\ep$. To make the total action $I^{(tot)}_{E}$ finite,
we need to add the counter terms
\be
I^{(c.t.)}_E=\f{L}{8\pi G_N}\int_{M}\s{\gamma}+\f{L}{8\pi G_N}\int_{P}\s{\Sigma}- \f{L}{16\pi G_N}\cdot \log \ep\cdot\left(\int_M\s{\gamma}R+2\int_{P}\s{\Sigma}K\right).
\ee

This leads to
\be
I^{(tot)}_{E}=-\f{\eta_*}{4G_N}-\f{L}{4G_N}-\f{L}{4G_N}\log r_B. \label{finth}
\ee
Therefore, the boundary entropy $S_{bdy}$, which is defined by the finite contribution to $-I^{(tot)}_{E}$ in the presence of the boundary $P$,  is given by
\be
S_{bdy}=\f{\eta_*}{4G_N}. \label{bent}
\ee
This is the same as the conclusion in \cite{Ta,Fujita:2011fp}, where $I^{(bdy)}_{E}$ was not taken into account. Indeed, (\ref{bent}) agrees with another calculation of $S_{bdy}$ using the holographic entanglement entropy \cite{RT,Takayanagi:2012kg}. Notice also that the logarithmic term in (\ref{finth}), which is proportional to the Weyl anomaly, is not affected by the new term $I^{(bdy)}_{E}$. In section \ref{bcfttwo}, we will generalize the calculation of the logarithmic
term to the case where $P$ is an arbitrary closed loop.

In this way, most of physical quantities do not change by the addition of the new boundary term $I^{(bdy)}_{E}$. However, there is at least one exception, which is the boundary energy momentum
tensor, as we will discuss in the next subsection.

\subsection{Holographic Boundary Energy Momentum Tensor}

In the general setups of AdS/CFT, a convenient choice of coordinate is known as the
Fefferman-Graham coordinate and is defined by
\be
ds^2=\f{L^2}{4\rho^2}d\rho^2+\f{1}{\rho}g_{ij}(x,\rho)dx^i dx^j. \label{FGC}
\ee
The special case $g_{ij}=\delta_{ij}$ corresponds to the pure AdS$_{d+1}$ and the
coordinate $\rho$ is related to $w$ in (\ref{pads}) via $\rho=w^2$. The AdS boundary
$M$ is situated at $\rho=0$ and thus the metric of $M$ in the gravity dual is given by $\gamma_{ij}=\lim_{\rho\to 0}\f{g_{ij}}{\rho}$.
The metric of $M$ in the BCFT$_{d}$ is given by $\lim_{\rho\to 0} g_{ij}=g_{ij}^{(0)}$.

The energy stress tensor $T_{ij}$ is a useful physical quantity to
characterize the property of CFTs in such general setups. It is defined by
the variation of the action $I_{CFT}$ with respect to the metric $g^{(0)}_{ij}$
\be
T_{ij}=-\f{2}{\s{g^{(0)}}}\cdot\f{\delta I_{CFT}}{\delta g^{(0)ij}}.
\ee

The holographic energy momentum tensor \cite{BaKr,MyS,Sk} is defined so that it is
 proportional to the derivative of the total gravity action with respect to the AdS boundary metric $\gamma_{ij}$ (called Brown-York tensor \cite{BrYo}):
\be
T^{(AdS)}_{ij}=\lim_{\rho\to 0}\left[\f{\rho^{1-\f{d}{2}}}{8\pi G_N}(K_{ij}-\gamma_{ij}K)+(\mbox{counter terms})\right]. \label{es}
\ee

Moreover, we would like to point out that in the AdS/BCFT setup we can also calculate
the boundary analogue of the energy momentum tensor $B_{\ap\beta}$, which has been first introduced in \cite{McOs} from a field theoretic viewpoint.
In BCFTs, this boundary energy momentum tensor is defined by taking the variation of the action $I_{BCFT}$ with respect to the metric $\sigma_{\ap\beta}$ on $\de M$
\be
B_{\ap\beta}=-\f{2}{\s{\sigma}}\cdot\f{\delta I_{BCFT}}{\delta \sigma^{\ap\beta}}.
\ee
In the gravity side, we argue the following holographic formula by taking the derivative of (\ref{boundr}) with respect to $\Sigma_{\ap\beta}$ (notice the relation
$\lim_{\rho\to 0}\rho\cdot\Sigma_{\ap\beta}=\sigma_{\ap\beta}$)
 \be
 B^{(AdS)}_{\ap\beta}=\lim_{\rho\to 0}\left[\f{\rho^{\f{3}{2}-\f{d}{2}}}{8\pi G_N}(2\theta-\pi)\Sigma_{\ap\beta}+(\mbox{counter terms})\right]. \label{bes}
 \ee
We will later evaluate $B^{(AdS)}_{\ap\beta}$ explicitly in AdS$_3/$BCFT$_2$ and confirm that it plays the crucial role on the consistency with the Weyl anomaly.

\section{AdS$_3/$BCFT$_2$ with Arbitrary Boundaries and Conformal Anomaly} \label{bcfttwo}

In previous examples, the AdS/BCFT has been constructed when the boundary $\de M$ of the BCFT takes special shapes such as hyperplanes or round spheres. Therefore we would like to generalize the AdS/BCFT construction and analyze the cases where $\de M$ take arbitrary shapes. In this section we will employ the Fefferman-Graham coordinate (\ref{FGC}) and mainly focus on the AdS$_{3}/$BCFT$_2$.

In the near AdS boundary limit, we can expand \cite{HeSk,Sk}
\be
g_{ij}=g^{(0)}_{ij}+\rho g^{(2)}_{ij}+\rho\log\rho\ h^{(2)}_{ij}+\ddd.
\ee

The profile of the boundary $Q$ in the AdS$_3$ is described by the constraint (setting
$x^1=x$ and $x^2=y$)
\be
x=x(y,\rho),
\ee
which is expanded as
\be
x(y,\rho)=x^{(0)}(y)+\s{\rho}\ x^{(1)}(y)+\rho\ x^{(2)}(y)+\ddd.
\ee

\subsection{Einstein Equation}
The Einstein equation in the $d+1$ dimensional Fefferman-Graham coordinate (\ref{FGC})
can be summarized as follows \cite{HeSk}
\ba
&& \rho(2g''_{ij}-2g'_{ik}g^{kl}g'_{lj}+g^{kl}g'_{kl}g'_{ij})-L^2R(g)_{ij}
-(d-2)g'_{ij}-g^{kl}g'_{kl}g_{ij}=0,\no
&& g^{jk}(\nabla_{i}g'_{jk}-\nabla_{k}g'_{ij})=0,\no
&& g^{ij}g''_{ij}-\f{1}{2}g^{ij}g'_{jk}g^{kl}g'_{li}=0, \label{einq}
\ea
where $R(g)_{ij}$ is the $d$ dimensional Ricci tensor for the metric $g_{ij}$, regarding
$\rho$ as a constant.

In the $d=2$ case, by expanding the Einstein equations (\ref{einq}) about the powers of $\rho$,
we obtain \ba && h^{(2)}_{ij}=0,\no &&
g^{(0)ij}g^{(2)}_{ij}=-\f{L^2}{2}R^{(0)}, \label{anomg} \ea
where $R^{(0)}_{ij}$ is the Ricci tensor for $g^{(0)}_{ij}$.
Note that $g^{(2)}_{ij}$ is not completely fixed and this ambiguity, for example, leads to black hole solutions with various temperatures.

\subsection{Boundary Condition}

Next we would like to solve the boundary condition (\ref{eqbein}).
We proceed by assuming that the boundary metric $g^{(0)}$ is flat
\be
g^{(0)}_{ij}=\delta_{ij}. \label{fby}
\ee
In the leading order of $\rho$ expansion, (\ref{eqbein}) leads to
\be
x^{(1)}(y)=\f{TL^2\s{1+(\de_y x^{(0)})^2}}{\s{1-L^2T^2}}. \label{x1}
\ee

In the next order, we find
\ba
x^{(2)}(y)=\f{L^2\left(1+L^2T^2(\de_y x^{(0)})^2\right)
(\de_y^2 x^{(0)})}{2(1-L^2T^2)(1+(\de_y x^{(0)})^2)}. \label{x2}
\ea

It may be useful to consider the solutions with the Lorentzian signature so that they describe holographic time-dependent backgrounds. For this, we can wick rotate the $x$ coordinates as
$x=it$. This leads to the following solutions instead of (\ref{x1}) and (\ref{x2}):
\be
t^{(1)}(y)=\f{TL^2\s{(\de_y t^{(0)})^2-1}}{\s{1-L^2T^2}}, \ \ \ \ \
t^{(2)}(y)=\f{L^2\left(1-L^2T^2(\de_y t^{(0)})^2\right)
(\de_y^2 t^{(0)})}{2(1-L^2T^2)(1-(\de_y t^{(0)})^2)}.
\ee

\subsection{Partition Function}

Now we would like to evaluate the Euclidean partition function (\ref{totac}). Since we are interested especially in the logarithmically divergent term, we need to evaluate the main part
$I_E$, which can be simplified as follows
\be
I_E=\f{1}{4\pi L^2G_N}\int_N \s{g}-\f{T}{8\pi G_N}\int_{Q}\s{h}.
\ee

We can expand $I_E$ with respect to $\rho$ using the formula such as
\be
\s{g}=\f{L}{2\rho^2}\left(1+\f{\rho}{2}\mbox{Tr}[g_{(0)}^{-1}g^{(2)}]+\ddd\right).
\ee

The boundary $Q$ is described by a closed loop described by $x=x(y,\rho)$ and it is assumed to have two branches for a fixed
$y$ and $\rho$, which are denoted by $x_{+}(y,\rho)$ and $x_{-}(y,\rho)$ such that
we always have $x_{+}(y,\rho)>x_{-}(y,\rho)$. The region inside $Q$ is given by
$x_{-}(y,\rho)<x<x_{+}(y,\rho)$. We define $\Delta x(y,\rho)=x_{+}(y,\rho)-x_{-}(y,\rho)$.

In the end we can evaluate the logarithmically divergent term in $I_E$ by introducing the
UV cut off as $\rho>\ep^2$:
\ba
S_E&=&\f{1}{4\pi L^2G_N}\left[-\f{L}{2}\int dy \Delta x^{(2)}(y)\cdot \log \ep^2\right]\no
&&+\f{T}{8\pi G_N}\cdot \log\ep^2\int dy\ \Delta\left[\f{2x^{(1)}x^{(2)}+L^2(\de_y x^{(0)})(\de_y x^{(1)})}{2\s{L^2+(x^{(1)})^2+L^2(\de_y x^{(0)})^2}}\right]\no
&& =-\f{L}{4\pi G_N}\cdot
\log \ep \cdot \int dy \Delta \left(\f{\de_y^2x^{(0)}}{2(1+(\de_y x^{(0)})^2}\right).
\ea

Notice that the last term is topological because
\be
\int dy \Delta \left(\f{\de_y^2x^{(0)}}{2(1+(\de_y x^{(0)})^2}\right)
=\Delta\left[\f{1}{2}\arctan(\de_y x^{(0)})\right].
\ee

By extending this result to curved spaces using (\ref{anomg}),  we finally obtain
\ba
I_E&=&\log \ep\cdot \f{L}{16\pi G_N}\cdot
\left(\int_M \s{g^{(0)}}R^{(0)}+2\int_{\de M}\s{h^{(0)}}K^{(0)}\right)\no
&=& \f{c}{6}\chi(M)\cdot \log \ep, \label{anom}
\ea
where where we employed the well-known relation $c=\f{3L}{2G_N}$ \cite{BrHe}; $\chi(M)$ is the Euler number of $M$; $K^{(0)}$ is the trace of extrinsic curvature of
the curve $x=x^{(0)}(y)$, given by
\be
K^{(0)}=-\f{\de_y^2x^{(0)}(y)}{(1+(\de_y x^{(0)}(y))^2)^{3/2}}.
\ee

In this way we nicely reproduce the logarithmic term in the BCFT partition function,
which is expected from the Weyl anomaly.

\subsection{Analysis of Boundary Energy Momentum Tensor}

The trace of the holographic (bulk) energy momentum tensor (\ref{es}) for
the flat space BCFT (\ref{fby}) becomes trivial in our setup
\be
g^{(0)ij}T^{(AdS)}_{ij}=0, \label{bem}
\ee
as follows from (\ref{anomg}). One may immediately wonder if
this may contradict with the fact that the logarithmic term (\ref{anom}) shows a non-vanishing trace anomaly. However, this is not the case if we take into account the boundary energy momentum
tensor $B_{\ap\beta}$. Using the holographic formula (\ref{bes}) we can evaluate as follows
\be
B^{(AdS)}_{yy}=\f{1}{8\pi G_N}\lim_{\rho\to 0}\left[\s{\rho}\left(2\theta-\pi
+\arccos T\right)\Sigma_{yy}\right],
\ee
where the term proportional to $\arccos T$ is the counter term.
By using the inner product of the two unit normal vectors at the boundary $P=\de M=\de Q$:
\be
n_M\cdot n_Q=T+\f{x^{(0)''}(y)\s{1-T^2}}{(1+(x^{(0)'}(y))^2)^{3/2}}\s{\rho}+O(\rho),
\ee
finally we obtain
\be
B^{y(AdS)}_{y}=\f{1}{8\pi G_N}\f{x^{(0)''}(y)}{(1+(x^{(0)'}(y))^2)^{3/2}}=-\f{c}{12\pi}K^{(0)}. \label{Byy}
\ee

We can confirm that the total Weyl anomaly is consistent with (\ref{Byy}) as follows. The variation
of the gravity action is given by
\be
\delta I_E=-\f{1}{2}\int_M \s{g}T_{ij}\delta g^{(0)ij}-\f{1}{2}\int_{\de M}\s{\sigma}
B_{\ap\beta}\delta \sigma^{\ap\beta},
\ee
where note that $\sigma_{\ap\beta}$ is the same as $h^{(0)}_{\ap\beta}$ in (\ref{anom}).
For the infinitesimal Weyl transformation $\delta g_{ij}=2\ep g_{ij}$ and $\delta \sigma_{\ap\beta}=2\ep \sigma_{\ap\beta}$, we find
\be
\delta_\ep I_E=\ep\left[\int_M \s{g}T^i_i+\int_{\de M}\s{\sigma}
B^i_{i}\right]=-\f{\ep c}{6}\chi(M). \label{Weap}
\ee
This agrees with the logarithmic term in (\ref{anom}), which satisfies
\be
r_B\f{\de I_E}{\de r_B}=-\f{c}{6}\chi(M).
\ee

\subsection{Analysis in Higher Dimensions} \label{fgh}

One may think that we can generalize this analysis in higher dimensions $d>2$ . However,
this is not the case as we will see below. For example, consider $d=3$ case, where
we can expand the metric as
\be
g_{ij}=g^{(0)}_{ij}+\s{\rho}g^{(1)}_{ij}+\rho g^{(2)}_{ij}+\rho^{3/2}\ g^{(3)}_{ij}+\ddd. \label{expro}
\ee
and  the Einstein equation (\ref{einq}) can be solved as
\ba
&& g^{(1)}_{ij}=0, \no
&& g^{(2)}_{ij}=L^2\left(R_{ij}-\f{1}{4}Rg_{ij}\right), \no
&& g^{(0)ij}g^{(3)}_{ij}=0.
\ea
The profile of $Q$ can be specified by
\be
x=x(y,z,\rho).
\ee

We can analyze the boundary Einstein equation order by order in the $\rho$ expansion.
The leading order relation determines $x^{(1)}$. However, the second order equations
lead to the constraints
\ba
&& \de_y^2x^{(0)}\left(1+(\de_z x^{(0)})^2\right)=\de_z^2x^{(0)}\left(1+(\de_y x^{(0)})^2\right), \no
&& \left(1+(\de_z x^{(0)})^2\right)\de_y\de_z x^{(0)}=(\de_y x^{(0)})(\de_z x^{(0)})(\de_z^2 x^{(0)}).
\ea
This does not have any solutions\footnote{To see this quickly, we can assume $x^{(0)}$ is infinitesimally small and then the linearized equations are
$\de_y^2x^{(0)}-\de_z^2x^{(0)}=0$ and $\de_y\de_z x^{(0)}=0$. They allow
only solutions which are linear or quadratic with respect to $y$ and $z$.} except when the boundary $Q$ is given by
planes or spheres, which are already known solutions as reviewed in section \ref{examples}.

One may think that this shows that we cannot construct any gravity solutions dual a BCFT on $M$ for generic choice of the boundary $\de M$. This is clearly paradoxical because the BCFT side is well-defined for any $\de M$, though the generic choice of $\de M$ breaks the $SO(2,d-1)$ boundary conformal invariance. We will resolve this puzzle in the section (\ref{adsf}) soon later. The upshot is that the $\rho$ expansion (\ref{expro}) breaks down at the boundary $Q$ and that we need to employ a different coordinate system.

\section{AdS$_4/$BCFT$_3$ with Arbitrary Boundaries} \label{adsf}

Consider the AdS$_4$/BCFT$_{3}$ setup with the three dimensional
boundary $Q$. We can choose the Gaussian normal coordinate (\ref{Gauss}), where
$Q$ is situated at $\eta=\eta_*$ and $N$ is extended in the region
$-\infty<\eta<\eta_{*}$. The extrinsic curvature is given by (\ref{formulak}).
The (vacuum) Einstein equation is decomposed into the constraints \ba &&
R^{(3)}+K^2-K_{ab}K^{ab}=2\Lambda\left(=-6/R^2\right),
\label{hc} \no && \nabla^a(K_{ab}-h_{ab}\cdot K)=0, \label{mc} \ea
and evolution equations of $K_{ab}$ with respect to $\eta$.

If we consider the boundary matter field, the boundary condition
takes the general form (\ref{bein}). The constraint (\ref{mc})
is equivalent to the conservation of
boundary energy-momentum tensor $T^{Q}_{ab}$ \be \nabla^a
T^{Q}_{ab}=0. \label{cen} \ee On the other hand, (\ref{hc}) can be
expressed as \be
R^{(3)}-(T^{Q}_{ab})^2+\f{1}{2}(T^{Q})^2=2\Lambda. \label{ncs}
\ee

Thus for any matter stress tensor $T^{Q}_{ab}$ which satisfies the
conservation (\ref{cen}) and a constraint (\ref{ncs}), we can always
construct a bulk metric by (numerically) solving the Einstein
equation without any obstruction.

\subsection{Construction of Perturbative Solutions}

We would like to construct perturbative solutions. We will set the AdS radius
to be one $L=1$ in this section just for the simplification. We express
the metric in (\ref{Gauss}) as follows \ba
h_{ab}(\eta,x,y,z)=\f{\cosh^2\eta}{z^2}\delta_{ab}+\delta h_{ab}(\eta,x,y,z), \ea
choosing the coordinates
$(u^1,u^2,u^3)=(x,y,z)$. We treat $\delta h_{ab}$ as a perturbation and keep only
 the first order. Notice that the unperturbed four dimensional metric is the same as (\ref{metads}) and
thus coincides with the pure AdS$_4$ (\ref{pads}) via the coordinate transformation
(\ref{core}).

Now, what we need to do is to solve the Einstein equation with the boundary condition
\be
\f{\de h_{ab}(\eta_*,x,y,z)}{\de \eta}=T\cdot h_{ab}(\eta_*,x,y,z). \label{bck}
\ee
as follows from (\ref{eqbein}). Notice that the tension is related to $\eta_*$ via
(\ref{tension}), which is given by $T=2\tanh \eta_*$ in the current setup.

The Einstein equation can be decomposed into the constraints (\ref{mc})
and the $ab$ components of the Einstein equation
\be
R_{ab}-\f{1}{2}g_{ab}R+\Lambda g_{ab}=0.  \label{Eab}
\ee

We are interested in the metric perturbations $\delta h_{ab}$ which depends on $x$ and $y$. We only work on the linear order of this perturbation theory and neglect higher orders.
We will perform the Fourier transformation with respect to $x$ and $y$. By employing the rotation on the $x-y$ plane, we can set the wave vector for $y$ to be vanishing. Therefore we only consider the perturbation proportional to
$e^{ikx}$:
\be
\delta h_{ab}(\eta,x,y,z)=\delta h_{ab}(\eta,z,k)\cdot e^{ikx}.
\ee

Moreover, to solve the Einstein equation analytically, we also assume perturbative expansions
with respect to the coordinate $z$ near the AdS$_3$ boundary $z=0$. Then the perturbative solution to the three dimensional part of the Einstein equation (\ref{Eab}) with the boundary condition (\ref{bck}) can be found to be
\ba
&& h_{xx}(\eta,z,k)=a_{xx}\f{\cosh^2\eta}{z^2}+\f{b_{xx}(\eta)}{z}+O(1),\no
&& h_{xy}(\eta,z,k)=a_{xy}\f{\cosh^2\eta}{z^2}+\f{b_{xy}(\eta)}{z}+O(1),\no
&& h_{yy}(\eta,z,k)=a_{yy}\f{\cosh^2\eta}{z^2}+\f{b_{yy}(\eta)}{z}+O(1),\no
&& h_{xz}(\eta,z,k)=a_{xz}\f{\cosh^2\eta}{z^2}+\f{b_{xz}(\eta)}{z}+O(1),\no
&& h_{yz}(\eta,z,k)=a_{yz}\f{\cosh^2\eta}{z^2}+\f{b_{yz}(\eta)}{z}+O(1),\no
&& h_{zz}(\eta,z,k)=\ \ \ \ \ \ \ 0  \ \ \ \ \ + \   \f{b_{zz}(\eta)}{z}+O(1),  \label{hexp}
\ea
where $a_{ab}$ are arbitrary (small) constants. The functions $b_{ab}(\eta)$ are defined by
\ba
&& b_{xx}(\eta)=-2(q_1-ia_{xz}k)\cosh^2\eta-q_2 \cdot \cosh\eta\left(1+\left(2(\arctan(e^{\eta})-\arctan(e^{\eta_*})\right)\cdot\sinh\eta\right),\no
&& b_{xy}(\eta)=ia_{yz}k\cosh^2\eta+\f{q_3\cosh\eta}{\arctan(e^{\eta_*})}
\left(-2+4\left(\arctan(e^{\eta_*})-\arctan(e^{\eta})\right)\sinh\eta\right)   ,\no
&& b_{yy}(\eta)=-2q_1\cosh^2\eta+q_2\cosh\eta
\left(1+2\left(\arctan(e^{\eta})-\arctan(e^{\eta_*})\right)\sinh\eta\right)         ,\no
&& b_{xz}(\eta)=q_4\cosh^2\eta    ,\no
&& b_{yz}(\eta)=q_5\cosh^2\eta     ,\no
&& b_{zz}(\eta)=2q_1\cosh^2\eta,
\ea
where $q_1,q_2,\ddd,q_5$ are arbitrary (small) constants.
Finally we can also confirm that these solutions satisfy the constraints (\ref{mc}).
In this way, we can construct perturbative solutions with several parameters.

\subsection{Analysis of Explicit Solutions}

To find a simplest non-trivial solution we would like to set
\ba
&& a_{xx}=a_{xy}=a_{yy}=a_{yz}=q_3=q_4=q_5=0,  \label{casse}
\ea
with $a_{xz}$, $q_1$ and $q_2$ chosen to be arbitrary.

For large $\eta$ limit (i.e. AdS boundary limit $\xi >>w$), they behave as
\ba
&& b_{xx}(\eta)\simeq \left(2ia_{xz}k-2q_1+(2\arctan(e^{\eta_*})-\pi)q_2\right)\f{\xi^2}{w^2},\no
&& b_{yy}(\eta)\simeq \left(-2q_1+(\pi-2\arctan(e^{\eta_*}))q_2\right)\f{\xi^2}{w^2},\no
&& b_{zz}(\eta)\simeq 2q_1\cdot\f{\xi^2}{w^2}, \no
&& b_{xy}(\eta) \simeq b_{xz}(\eta)\simeq b_{yz}(\eta)\simeq 0, \label{relazz}
\ea
where notice that $e^\eta\simeq \f{2\xi}{w}$ in the limit $\eta\to\infty$.

We assume that $q_1$ and $q_2$ are imaginary and define
\ba
&& \beta_{xx}=2a_{xz}k-2|q_1|+(2\arctan(e^{\eta_*})-\pi)|q_2|,\no
&& \beta_{yy}=-2|q_1|+(\pi-2\arctan(e^{\eta_*}))|q_2|,\no
&& \beta_{zz}=2|q_1|. \label{betapp}
\ea

By using the relation (\ref{core}) or equally
\be
z=\s{w^2+\xi^2},\ \ \ \sinh\eta=\f{\xi}{w},
\ee
the total metric can be expressed as
\ba
ds^2&=&\f{1}{w^2}\Biggl[\!dw^2\!+\!d\xi^2\!+\!dx^2\!+\!dy^2\!+\!
\f{2a_{xz}\cos(kx)}{\s{w^2+\xi^2}}(wdw+\xi d\xi)dx \no
&&  -\f{\beta_{zz}\sin(kx)}{\s{w^2+\xi^2}}(wdw+\xi d\xi)^2
-\f{w^2\sin(kx)|b_{xx}(\eta)|}{z}dx^2-\f{w^2\sin(kx)|b_{yy}(\eta)|}{z}dy^2\Biggr].
\ea
We perform the coordinate transformation
\ba
&& x\to x-a_{xz}\cos(kx)\s{w^2+\xi^2}.\no
\ea
Then the metric is rewritten as follows up to the linear order of the perturbation
\ba
ds^2&=&\f{1}{w^2}\Biggl[dw^2+d\xi^2+dy^2+\left(1+2a_{xz}k\sin(kx)\s{w^2+\xi^2}\right)
dx^2 \no
&&  -\f{\beta_{zz}\sin(kx)}{\s{w^2+\xi^2}}(wdw+\xi d\xi)^2
-\f{w^2\sin(kx)|b_{xx}(\eta)|}{z}dx^2-\f{w^2\sin(kx)|b_{yy}(\eta)|}{z}dy^2\Biggr].\label{metpt}
\ea

\subsection{Relation to Fefferman-Graham Coordinate}

We would like to rewrite the metric (\ref{metpt}) in terms of Fefferman-Graham coordinate so that we can analyze the structure of AdS boundary. For this purpose we perform the coordinate transformation
\be
\hat{w}=w-\Delta w(w,\xi,x),\ \ \ \hat{\xi}=\xi+\Delta \xi(w,\xi,x),
\ee
where
\ba
&& \Delta w(w,\xi,x)=\f{\beta_{zz}\sin(kx)}{4}\left[w\s{w^2+\xi^2}-\xi^2\log(2(w+\s{w^2+\xi^2}))\right],\no
&&  \Delta \xi(w,\xi,x)=\f{\beta_{zz}\sin(kx)}{4}\left[w\xi+2w\xi\log(2(w+\s{w^2+\xi^2}))\right].
\ea
Then we find the metric takes the following Fefferman-Graham form in the new coordinate $\hat{w},\hat{\xi}$ (we omit the symbol $\ \hat{}\ $ below):
\ba
ds^2=\f{dw^2}{w^2}+\f{(1+\Delta g_{\xi\xi})d\xi^2+(1+\Delta g_{xx})dx^2+(1+\Delta g_{yy})dy^2}{w^2}.\label{FGmet}
\ea
Here we defined
\ba
&& \Delta g_{\xi\xi}=\f{\beta_{zz}\sin(kx)}{2}\left[\f{w}{2}+\f{\xi^2 w}{w^2+\xi^2+w\s{w^2+\xi^2}}-\f{2\xi^2}{\s{w^2+\xi^2}}+w\log(2(w+\s{w^2+\xi^2}))\right],\no
&& \Delta g_{xx}=\sin(kx)\left[2a_{xz}k\s{w^2+\xi^2}-\f{w^2}{\s{w^2+\xi^2}}|b_{xx}(\eta)|\right],\no
&& \Delta g_{yy}=-\sin(kx)\f{w^2}{\s{w^2+\xi^2}}|b_{yy}(\eta)|,
\ea
where we neglect terms with higher powers of $w$ and $\xi$ than the ones included in the above.
Notice that our perturbative solution (\ref{FGmet}) is only valid when $\xi$ and $w$ are
small.

We define the three dimensional part of the metric $ds^2_{(3)}$ from this metric (\ref{FGmet}) as
\ba
ds^2= \f{dw^2+ds_{(3)}^2}{w^2}.
\ea
The metric of the AdS boundary is obtained from $ds^2_{(3)}$  by taking the limit $w\to 0$
\ba
&& ds_{(3)}^2\Bigl|_{w=0}= \left(1+(2a_{xz}k-\beta_{xx})\sin(kx)\xi\right)dx^2
+\left(1-\beta_{yy}\sin(kx)\xi\right)dy^2\no
&& \ \ \ \ \ \ \ \  +\left(1-\beta_{zz}\sin(kx)\xi\right)d\xi^2,
\ea
where we employed (\ref{relazz}).

Now we would like to concentrate on the case where the AdS boundary becomes flat.
 We find that this corresponds to the case
 \be
\beta_{yy}=-2|q_1|+(\pi-2\arctan(e^{\eta_*}))|q_2|=0.
\ee
 Indeed, in this case, if we further perform the coordinate transformation
\ba
&& x=\ti{x}+\f{(2a_{xz}k-\beta_{xx})\cos(k\ti{x})}{2k}\ti{\xi},\no
&& \xi=\ti{\xi}-\f{\sin(k\ti{x})}{2}
\left(\f{2a_{xz}k-\beta_{xx}}{k^2}+\f{\beta_{zz}}{2}\ti{\xi}^2\right),
\ea
we find that the metric at the boundary $w=0$ becomes flat:
\be
ds^2\simeq \f{dw^2+d\ti{\xi}^2+d\ti{x}^2+dy^2}{w^2}.
\ee

On the other hand, the boundary $Q$, which is originally defined by $\xi=0$ is now described by
\be
\ti{\xi}=\f{(2a_{xz}k-\beta_{xx})\sin(k\ti{x})}{2k^2}=\f{2|q_1|}{k^2}\sin(k\ti{x}). \label{cbf}
\ee
Thus the boundary of the BCFT is now a curved surface.

As we have seen in our construction of perturbative solutions, we encounter expansion in terms of $z=\s{w^2+\xi^2}$. Thus the usual Fefferman-Graham coordinate with the $w=\s{\rho}$ expansion breaks down at the boundary $Q$ i.e. $\xi=0$. This was the reason why we could not find solutions with
curved BCFT boundaries via the Fefferman-Graham expansion in section \ref{fgh}.

\section{Towards Higher Dimensional g-Theorems}

\begin{figure}
   \begin{center}
     \includegraphics[height=4cm]{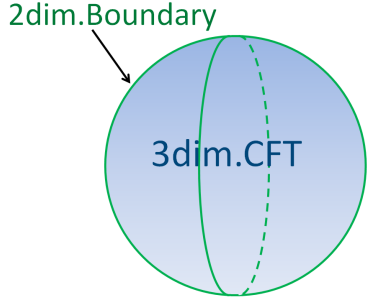}
   \end{center}
   \caption{The setup of BCFT$_3$ to calculate the boundary central charge $c_{bdy}$.}\label{fig:bth}
\end{figure}

In two dimensional CFTs, there is a famous Zamolodchikov's c-theorem \cite{cth},
which argues that the
central charges in CFTs are decreased under RG-flow. Moreover, we can construct so called c-function which
is monotonically decreasing under the RG-flow and which coincides with the central charges at the fixed points.
It is natural to expect something similar in a BCFT$_3$ i.e. a CFT on a three dimensional manifold $M$ with a boundary $\de M$. Here we consider the boundary RG flow, while the three dimensional bulk is kept conformally invariant. We assume that $M$ is given by the three dimensional Euclidean round ball $B_3$ with radius $r_B$. Its boundary $\de M$ is a round $S^2$ with the same radius. See Fig.\ref{fig:bth} for this setup. This radius $r_B$ can be regarded as the length scale under the RG flow.
In this setup, the idea of the boundary central charge $c_{bdy}$ was introduced in
\cite{Fujita:2011fp} and there a holographic analysis based on AdS/BCFT showed that this quantity, extended
 to non-conformal theories in an appropriate way, is a monotonically decreasing function under the RG flow. The boundary central charge (or c-function) $c_{bdy}$ at the length scale $r_B$ is defined by
\be
c_{bdy}(r_B)=3r_B\f{d\log Z_{BCFT_3}}{dr_B}\left(=-3r_B\f{d I_{BCFT_3}}{dr_B}\right), \label{defc}
\ee
in terms of the derivative of the partition function $Z_{BCFT_3}$ of the BCFT$_3$ on $M$.
The normalization of $c_{bdy}$ is chosen such that this agrees with that of the standard central charge in two dimensional CFTs if the bulk theory is completely decoupled from the boundary. If we consider the bare partition function $Z_0$, at the fixed point, $c_{bdy}$ is the coefficient of the logarithmically divergent term
\be
\log Z_0=\mbox{power divergences}+ \f{c_{bdy}}{3}\log \f{r_B}{a}+\mbox{finite terms}.
\ee

We conjecture that in any quantum field theories $c_{bdy}$ (\ref{defc})
satisfies the monotonicity property
\be
\f{dc_{bdy}(r_B)}{dr_B}\leq 0. \label{bthm}
\ee
This can be regarded as a three dimensional version of the g-theorem \cite{AfLu}.
 An analogue of c-theorem for three dimensional CFTs without boundaries has already been formulated in \cite{JKPS,Klebanov:2011gs,CaHu} and is called F-theorem. It is straightforward to generalize (\ref{bthm}) to
 much higher dimensions as long as $d$ is odd. By comparing with the c-theorems in higher dimensions
 \cite{Cardy:1988cwa,hcth,KoSc}, we can conjecture
 \be
 (-1)^{\f{d+1}{2}}r_B\f{d\log Z_{BCFT_d}}{dr_B}\leq 0. \label{bthmg}
 \ee
 Below we will give a few supporting evidences of these higher dimensional g-theorems.

\subsection{Perturbative Confirmation}

We consider the case where the BCFT$_{d}$ on the round ball $B_{d}$ for $d$ odd. We perturb this theory by a boundary operator $\mathcal{O}(x)$
\be
S=S_{BCFT_{d}}+\lambda_0\int_{S^{d-1}} dx^{d-1}\s{G}\mathcal{O}(x),
\ee
where $\lambda_0$ is a bare coupling constant.
The boundary conformal dimension of $\mathcal{O}(x)$ is defined to be $\Delta=d-1-y$ and we assume the
relevant perturbation which is nearly marginal $0<y<<1$.

The boundary conformal invariance constrains the two and three point function of a boundary operator just as in the standard $d-1$ dimensional CFT:
\ba
&& \langle \mathcal{O}(x)\mathcal{O}(w)\rangle =\frac{1}{|x-w|^{2(d-1-y)}},\no
&& \langle \mathcal{O}(x)\mathcal{O}(w) \mathcal{O}(z)\rangle=\frac{C}{|x-w|^{d-1-y}|w-z|^{d-1-y}|z-x|^{d-1-y}},  \label{crl}
\ea
where $x,w$ and $z$ lives on the flat space $R^{d-1}$ and we need to perform an obvious conformal transformation to obtain the two and three point functions on the $S^{d-1}$.

We define the dimensionless coupling by $g(\mu)=\lambda(\mu)\cdot \mu^{-y}$, where
$\mu$ is the energy scale which we identify $\mu=\f{1}{2r_B}$. The initial condition is set by
$g(\mu_0)=\lambda_0  \mu_0^{-y}$, where $\mu_0$ is the UV cutoff.
The $\beta$-function for the renormalized coupling $g(\mu)$ is given by \cite{Cardy:1988cwa,Klebanov:2011gs}
\be
\beta(g)\equiv \mu\f{d g(\mu)}{d \mu}=-yg+\f{\pi^{\f{d-1}{2}}}{\Gamma(\f{d-1}{2})} Cg^2+O(g^3). \label{befu}
\ee
By integrating this we obtain
\be
\lambda_0(2r_B)^y=g+\f{C\pi^{\f{d-1}{2}}}{ y \Gamma(\f{d-1}{2})}g^2+O(g^3).  \label{lze}
\ee

Now we evaluate the partition function. Using the results in \cite{Klebanov:2011gs}, we finally
obtain its perturbative expansion w.r.t $\lambda_0$, employing (\ref{crl}) and (\ref{lze}):
\ba
&& \log Z_{BCFT_d}=\f{\pi^{\f{d-1}{2}}}{2^{d-1}\Gamma (\f{d-1}{2})}\Biggl[\f{\lambda_0^2}{2} \cdot(2r_B)^{2y} \cdot \pi^{\f{d-1}{2}}\f{\Gamma(-\f{d-1}{2}+1)}{\Gamma(y)}\no
&&\ \ \ -\f{\lambda_0^3}{6}\cdot (2r_B)^{3y} \cdot \pi^{d-1}\f{\Gamma(\f{y}{2})^3 \Gamma(-\f{d-1}{2}+\f{3y}{2})}{\Gamma(y)^3 \Gamma(\f{d-1}{2})}\cdot C\Biggr]
+O(\lambda_0^4) \no
&& = \f{(-1)^{\f{d-1}{2}} \pi^{d-\f{1}{2}}}{\Gamma(\f{d-1}{2}+1)\Gamma(\f{d}{2})2^{d-2}}\left[\f{g^2}{2}+\f{1}{9y}
\f{\pi^{\f{d-1}{2}}}{\Gamma(\f{d-1}{2})}Cg^3 \right] +O(g^4),
\ea
where we keep only the leading term assuming $y$ is very small.

Up to this order we can find the boundary central charge as follows (remember $d$ is an odd integer)
\ba
&&  (-1)^{\f{d+1}{2}}r_B\f{d\log Z_{BCFT_d}}{dr_B}\no
&&=\f{\pi^{d-\f{1}{2}}}{\Gamma(\f{d-1}{2}+1)\Gamma(\f{d}{2})2^{d-2}} +
\left[g+\f{1}{3y}\f{\pi^{\f{d-1}{2}}}{\Gamma(\f{d-1}{2})}Cg^2 \right] \cdot \beta(g)+O(g^4) \no
&& =\f{ \pi^{d-\f{1}{2}}}{\Gamma(\f{d-1}{2}+1)\Gamma(\f{d}{2})2^{d-2}}
\left[-yg^2+\f{2}{3}\f{\pi^{\f{d-1}{2}}}{\Gamma(\f{d-1}{2})}Cg^3 \right] +O(g^4).
\ea
Finally, by taking the derivative of $r_B$ we obtain
\ba
 (-1)^{\f{d+1}{2}}r_B\f{d\log Z_{BCFT_d}}{dr_B}&=&-\f{ \pi^{d-\f{1}{2}}}{\Gamma(\f{d-1}{2}+1)\Gamma(\f{d}{2})2^{d-2}}\left[-2yg+2\f{\pi^{\f{d-1}{2}}}{\Gamma (\f{d-1}{2})}Cg^2 \right] \beta(g) +O(g^4) \no
&=& -\f{2\pi^{d-\f{1}{2}}}{\Gamma(\f{d-1}{2}+1)\Gamma(\f{d}{2})2^{d-2}}\beta(g)^2 +O(g^4).
\ea
Therefore we manage to show the property (\ref{bthmg}) in this perturbation theory.

Notice that in this argument the dynamics of the bulk conformal field theory is not relevant and
this proof is essentially reduced to that of the perturbative proof of c-theorem \cite{Cardy:1988cwa}.

\subsection{An Explicit Example: Massless Scalar Fields in BCFT$_3$}

We would like to evaluate contributions of boundary degrees of freedom to 1 loop partition functions of three
dimensional scalar fields, which become exact for free scalar fields. Since we are interested in CFTs in the bulk, we assume that they are massless scalars in three dimension.

In general, the one-loop partition function is expressed as
\be
Z=\int D\phi \exp \left( - \int _{\mathcal{M}} \phi^{i}\Delta_{ij}\phi^{j} \right).
\ee
We assume $\{\phi^{i} \}\ \ (i=1,2,\ddd,n)$ is a $n$ component bosonic field. $\Delta_{ij}$ is a second order
differential operator on $\mathcal{M}$ and one can consider the heat kernel $\hat{K}^{ij}(x,y;s)$
of the operator (we will closely follow \cite{Vreview})
\be
\Delta_{ij}\hat{K}_{jl}(x,y;s) =\f{\partial}{\partial s } \hat{K}_{il}(x,y;s).
\ee
The logarithm of the partition function in three dimension is given by
\be
\log Z_{BCFT_3}=\f{1}{2} \int ^{\infty}_{\epsilon^2} \f{ds}{s} \int_{\mathcal{M}} dx^3 \mbox{tr}\hat{K}(x,x;s).
\ee
where $\epsilon$ is UV cut off (lattice spacing). One can asymptotically expand the heat kernel near $s=0$

\be
\int_{\mathcal{M}} \mbox{tr}\hat{K}(x,x;s)=\f{a_{0}}{s^{\f{3}{2}}}+\f{a_{1}}{s}+\f{a_{2}}{s^{\f{1}{2}}}+a_{3} +\cdots ,
\ee
where $a_{i}$ s are heat kernel coefficients and can be written by geometric invariants such as various curvatures of $M$. The index $i$ denotes number of differentials they contain.
By substituting it to the partition function, we obtain
\be
\log Z_{BCFT_3}=\f{2}{3}\f{a_{0}}{\epsilon^3}+\f{a_{1}}{2\epsilon^2}+\f{a_{2}}{\epsilon}-a_{3} \log \epsilon  +\cdots.
\ee
In this way one can manifest  the divergent structure of the partition function.
When we consider a manifold without boundaries, $a_{j}$ vanish for all odd integers $j$ since all geometric invariants
contains even numbers of differentials and thus there are no log term in the partition function in three dimension. It is consistent with the fact that there is no trace anomaly in three
dimension. However, in three dimensional field theories with boundaries,
$a_{j}$ no longer vanishes for $j$ odd and there is a logarithmically divergent term.

 In general, we impose the following boundary condition at
 $\partial \mathcal{M}$ for $\{\phi^{i} \}$ of the form
 \begin{equation}
   \Pi_{-}\phi \textemdash _{\partial \mathcal{M}}=0,\hspace{2 cm} (\nabla_{n}+S) \Pi_{+}\phi \textemdash _{\partial \mathcal{M}}=0, \label{bcsc}
 \end{equation}
 where  $\Pi_{-}$ is hermitian projection operator of the $\phi$ such that $\Pi_{-}^2=1$ and  $\Pi_{+}=1-\Pi_{-}$. Notice that the $\Pi_{-}$ and $\Pi_+$ are the projections into the
 Dirichlet and (generalized) Neumann boundary conditions, respectively. The differential operator
 $\Delta_{ij}$ is chosen to be the Laplacian of $M$ for the massless scalars i.e. $\Delta_{ij}=-\delta_{ij}\cdot g^{ab}\nabla_a\nabla_b$. In this case, the heat kernel coefficient $a_{3}$ is given by the following formula as derived in \cite{BGV,Vreview}:
 \begin{eqnarray}
 a_{3}  &=& \f{1}{1536\pi}\int_{ \partial \mathcal{M}} \s{\sigma}
 dx^2\ {\rm tr}  \Bigl(16(\Pi_+-\Pi_-) R - 8(\Pi_+-\Pi_-) R^{\ap n}_{\ap n} \no
&& \ \ \ \ \ \ \ \  +(13\Pi _{+}-7\Pi _{-})K^2+(2\Pi _{+}+10\Pi _{-})K_{\ap\beta}K^{\ap\beta}+96SK+192S^2 \Bigr),\label{heatk}
 \end{eqnarray}
 where the trace is with respect to the index $i$ of $\phi^i$; $\sigma$ and $K_{\ap\beta}$ are the induced metric and the extrinsic curvature of $\de M$ in $M$, respectively; $R^{\ap n}_{\ap n}$ are components of curvature tensor in
 $M$ and $n$ represents the normal direction for $\de M$.

Now let us calculate the boundary central charges defined in (\ref{defc}). For this purpose we assume
the metric $M$ is flat and $\de M$ is a round $S^2$ with the radius $r_B$. Then the boundary central charge is given by $c_{bdy}=3a_{3}$. By using the formula (\ref{heatk}), for the Neumann and the Dirichlet boundary condition\footnote{The boundary condition (\ref{bcsc}) for non-vanishing $S$ breaks the boundary
conformal invariance and cannot be a fixed point of boundary RG flows. Therefore we only consider
$S=0$ i.e. the (purely) Neumann boundary condition here.}, we obtain the following boundary central charges:
\be
c_{bdy}(\mbox{Neumann})=\f{7}{16},\ \ \ \ c_{bdy}(\mbox{Dirichlet})=-\f{1}{16}.
\ee

Since it is clear that there is a RG flow from the Neumann to the Dirichlet just by adding the mass term at the boundary of the form
\be
\lambda\int_{\de M} dx^2  \phi^2,
\ee
the relation
\be
c_{bdy}(\mbox{Neumann})>c_{bdy}(\mbox{Dirichlet}),
\ee
 is consistent with our conjectured property (\ref{bthm}).

\section{Conclusions and Discussions}

In this paper, we studied the logarithmic terms in the partition functions of CFTs with boundaries (BCFTs)
 by employing both field theoretic and holographic approaches. In even dimensions, the coefficients of the log terms are related to the Weyl anomaly and thus the central charges. In odd dimensions, on the other hand, these coefficients lead to new quantities called boundary central charges $c_{bdy}$. A previous holographic analysis implies that $c_{bdy}$ are monotonically decreasing functions under the RG flows. This is interpreted as an odd dimensional analogue of the g-theorem known for two dimensional BCFTs. In this paper, we gave two evidences. One is that we showed this property in a leading order perturbation theory. The other is that we confirmed this in an explicit boundary
RG flow for massless scalar fields. These are purely based on the field theoretic calculations. It is certainly desirable to obtain an exact proof of this conjecture as well as various explicit examples.

We also did a related holographic analysis for BCFTs based on the AdS/BCFT formalism.
 In two dimensions, we gave an explicit holographic construction for an arbitrary shape of boundary and calculated its logarithmic term, confirming its consistency with the Weyl anomaly. We pointed out that we should add a codimension two boundary term in the gravity action, which has been missing so far in AdS/BCFT.
 This enables us to compute the energy momentum tensor $B_{\ap\beta}$ which is localized at the boundary.
 It is interesting to note that when a BCFT is defined on a round disk, the bulk energy momentum tensor is vanishing $T_{ij}=0$ because the gravity dual is given by a part of AdS$_3$. Our result shows that still this is consistent with the  Weyl anomaly. The reason why we have $T_{ij}=0$ is because we are considering the pure gravity theory where all solutions are locally AdS.  Therefore it is a very intriguing future problem to take into account back-reactions by considering a gravity theory coupled to various matter fields such as scalars or gauge fields so that the metric is no longer locally AdS.

We also gave perturbative solutions of gravity duals for the three dimensional BCFTs with any shapes of boundaries. We find that the standard Fefferman-Graham expansion
breaks down for generic choices of BCFT boundaries. It is another interesting future direction to
explore more on this AdS/BCFT in higher dimensions such as the construction of fully
back-reacted solutions and calculations of energy momentum tensors.

\section*{Acknowledgements}
TT is supported by JSPS Grant-in-Aid for Challenging
Exploratory Research No.24654057. TU is supported by JSPS
Research Fellowships for Young Scientists. 
A part of computations in this work were carried out at the Yukawa Institute Computer Facility. TT and TU are also
supported by World Premier International
Research Center Initiative (WPI Initiative) from the Japan Ministry
of Education, Culture, Sports, Science and Technology (MEXT).
MN would like to thank Kavli Institute for the Physics and Mathematics of the Universe for kind hospitality during his stay.
\appendix

\section{Gibbons-Hawking Term at Non-smooth Boundary}\label{ap:cusp}
Here we explain the extra boundary term \cite{Hay}
added to the standard Gibbons-Hawking term at a non-smooth boundary.
We consider a boundary specified by $x=x(y)$ in a three dimensional flat space
$ds^2=dx^2+dy^2+dz^2$.
Since the effect we are looking at is the one localized at the
 non-smooth points, this simple example captures all the essential parts.
 The unit normal vector reads
\be
n=(n_x,n_y,n_z)=\f{1}{\s{1+\f{1}{x'(y)^2}}}\left(-\f{1}{x'(y)},1,0\right).
\ee
 Then the Gibbons-Hawking term for the region $x_-\leq x \leq x_+$ is evaluated as
 \be
 \int \s{h}K=\int dy dz \f{x''(y)}{1+x'(y)^2}=\int dz \left[\arctan(x'(y))\right]^{x=x_+}_{x=x_-}.
 \ee
 Thus if the curve given by $y=-\tan\theta\cdot |x|$,  we find
 \be
 \int \s{h}K=\int dz (\pi-2\theta).
 \ee
 By covariantizing this expression, we finally obtain
 \be
 \int_{M\cup Q} \s{h}K=\int_{M'} \s{h}K+\int_{Q'} \s{h}K+\int_{M\cap Q} \s{\Sigma}(\pi-2\theta),
 \ee
 where $M\cap Q$ is where the cusps are located; $M'$ (and $Q'$) denote the points in
$M$ (and $Q$) except those in $ M\cap Q$. This reproduces (\ref{boundr}) for the Euclidean action (\ref{acte}).

\end{document}